# Examining the Relations between Household Saving Rate of Rural Areas in China and Migration

Fuhao Lou
University of Ottawa

Abstract. China has been developing very fast since the beginning of the 21$^{st}$ century. The net income of households has been increased a lot as well. Nonetheless, migration from rural areas to urban sectors tends to keep a high saving rate instead of consumption. This essay tries to use the conventional Ordinary Least Square regression, along with the method of Instrument Variable to test the problem of endogeneity, to discover the relationship between the saving rates of rural households and labor migration, controlling for other characteristic variables including having insurance, marital status, education, having children, health conditions. The assumption is that migration contributes positively to the dependent variable, meaning that migration could increase the household save rates. However, the conclusion is that it is negatively with the household save rates. All the other variables regarding education, health conditions, marital status, insurance, and number of children are negatively related with the household saving rates.

## I. INTRODUCTION

In China, the issues regarding the household saving rate and labor force have been brought up many times in different discussions of the economy. Chinese people, coming from different economic background, would like to save more than people living in other less developed countries. As the development of industrialization and urbanization, many rural workers migrate to urban sectors to seek higher-paying jobs. The revenue gained by them convert to higher saving rates instead of consumption. This essay tries to determine the relationship between rural household saving rates and migration.

Chinese economy has experienced many stages in terms of the transition since the end of 1970s. Before the formal reform of the economy in 1980s, the situations had been really hard that the government found that planned economy cannot be sufficient to create additional growth. Originally, the aim of the economy was designed for finishing and accelerating industrialization. For a long time, making sure the establishment of industrialization was the primary goal of the country. However, the international conditions for China to acquire the necessary capital to boost the process were hard. The economic and political block of the U.S. and western countries started in 1950. In 1960, because of the different political opinions from China's government and the government of the Soviet Union, the Soviet Union started to withdraw their engineering experts and different kinds of industrial supplies from China, making the process of China's industrialization worse. The first stage of industrialization is that different sectors, especially urban sectors need to accumulate enough capital to do massive productions. Given the harsh economic and political conditions in the period of 1950s to 1970s, the economic policy turned into planned economy. The structure of this planned economy is explained as a 'mock dual economy'[1](Christiansen 2006). Different sectors of governments bought the products from rural working force at a relatively low price and gained the revenue from these price differences to subsidy urban sectors related to heavy industrial products, expecting that the massive production of industrial products will recontribute to the economy and productivity of the rural sectors in the future.

  After the confirmation of beginning reforming the economy from this planned economy to the mixture of market economy and planned economy, many foreign investments flew into the coastal areas of China, setting up factories to produce daily products. On one hand, initially these companies and industries set up by foreign investors were mainly located in relatively urban sectors of the east coastal areas. Being a worker working in these factories can gain more salaries from working on agricultural productions in rural parts which were underdeveloped at that time. The difference of wage income attracts many residual workers migrate from rural to urban parts. On the other hand, these workers, after saving much money and transfer them back to their families, tend not to consume. Besides, because of many problems related to health care or insurance in underdeveloped rural parts, their families will tend to save than consume as well. Not only does the combination of the structural changes of the economy bring high GDP growth in large scale, but also it happens with low motivation of consumption, creating risks for the resilience of the growth of the economy in the future.

---

[1] It is constructed by using the residual revenues from trading with farmers to subsidy heavy industries.

   According to the Chinese Nation Bureau of Statistics, the discrepancies of the annual net income between rural and urban households become large over years. The expectation of the rural migration workers is that they want to reach the same level of income as urban households. Thus, this essay tries to use the conventional Ordinary Least Square regression, along with the method of Instrument Variable to test the problem of endogeneity, to discover the relationship between the saving rates of rural households and labor migration, controlling for other characteristic variables including having insurance, marital status, education, having children, health conditions. The assumption is that migration contributes positively to the dependent variable, meaning that migration could increase the household save rates. However, the conclusion is that it is negatively with the household save rates. All the other variables regarding education, health conditions, marital status, insurance, and number of children are negatively related with the household saving rates.

**II. LITERATURE REVIEW**

This essay is based on the paper called "The Influence of Rural Labor Migration on Household Saving Rate" written by the authors Yin Zhi-chao, Liu Tai-xing and Zhang Cheng (2019). The topic is related to labor and industrial economics which are very popular domains in the discussion of the development of modern China. It mainly tries to present the relations between rural household saving rates of Chinese households and rural labor migration from a complex and multidimensional background.

    On one hand, the development of China has been boosted since 1980s where China started to get the transfer of labor-intensive industries from western countries and some Asian countries like South Korea and Japan. The domestic market has been widened largely to attract foreign direct investment. On the other hand, the augmentation of investment and debt made by local governments increased the opening of jobs which were mainly related to manufacture and services. However, with the increasing of wage in mainly large cities and the policy of registration of residence, many rural citizens decided to migrate to large cities to find well-paid jobs, causing large migration in big cities. Because of the policy of residence, these people will not bring their families come together with them, instead, they tend to send money back to the households located in the rural parts of China. Hence, because of the money sent back and low cost of living in the rural areas, house saving rate among rural households keeps at a high level.

    Given the introduction of the background, this paper discusses the detailed problem of how the effect of labor migration from different rural areas impacts the household saving rate, along with trying to address the issue of endogeneity and heterogeneity. It used the data collected by CFPS (Chinese Family Panel Studies). It includes more than 14,000 households located in 25 different provinces in 2010. It also includes the detailed information about family income, occupation and extra. To investigate the research question, they cleaned the data and kept several important

variables, which are the number of migrants, household income and different kinds of expenditure and risks, by combing the characters of these households. In addition, in order to "solve the endogeneity issue caused by reverse causality and omitted variable bias, it chose the method of instrument variable"[2], meaning that they chose the instrument variable which is the ratio between labor force migration of the chosen family and that of the other families. The type of families picked by them are in the same region and from the same class. In addition, a problem that they encounter also is biased estimate due to sample selection bias. To adjust the biasness, this paper adopted two models. The first one is Treatment Effects Model, meaning that using Probit model to get into the household groups with migration, and then use the method of Ordinary Least Square to get the coefficients. The second model is to use PSM (Propensity Score Matching) to get the average effect of treatment.

In brief, the essay has found that, given all the outcomes of the experiments, household saving rate of rural families has been increased due to the migration of labor force; however, rural families tend to save the income instead of converting to consumption because of the potential scenarios which could reduce the income such as being sick. In addition, the migration of labor force reduced the gap of saving rate among rural families. Thus, it suggests the policy that local governments should try to increase the rate of expenditure of the rural families by stabilizing or reducing the potential risks.

Apart from the rural household rate, the relevant essay written by Chamon and Prasad (2008) sponsored by the International Monetary Fund studies the urban household rate from 1990 to 2005. Interestingly, although it is different from the first literature studying the rural household rate, the sameness of the results is that, according to the name of the essay, the aggregate (urban

---

[2] Two-stage Least Squares

4and rural) household saving rate has in fact risen by six percent points over the last decade. The data found by them is annual Urban Household Surveys conducted by China's National Bureau of Statistics.[3] It used a method of decomposition to estimate how saving rates are affected by different variables including age, time, and jobs, extra across different time periods. Interestingly, by controlling the demographic groups, they have found that people who are within younger and older age groups have the highest saving rates, meaning that they tend not to spend income into consumption. What let the author feel surprised is that households having higher saving rates are not the ones who have rapid income growth. The reasons of why this happen are somehow related from the result found in the last literature. Partially, the reasons are due to the expectations of the future expenditure of medical services and education expenses. Younger households may save more to prepare for the future study, and older households may worry about saving money for medical resources. Buying houses is another factor that causes saving rates grow. In addition, the other part found by the article is related to the structural change in the economy. At the end of 1990s, East Asian experienced financial crisis and the marketization of some national-owned Chinese companies began processing, the transition of the economy made people feel that they needed to save more money in order to get through difficult times. Not surprisingly, after conducting the experiments by controlling demographic variables over different time periods, they gave similar policy suggestions to increase the consumption and smooth the saving rates in the future. These policies are increasing the resources of public education, providing the medical supplies, and widening financial means for the households to invest and consume.

---

[3] "This is the first detailed examination of Chinese household saving behavior using micro data over a long span" (Chamon and Prasad 2008)

Besides the articles talking about the relations of household saving rates and Chinese households, Dynan, et.al (2004) tries to discover if the rich save more. It is an interesting and relevant question since after comparing the relations between saving rates of rural and urban households in China, the previous essays show that people having different economic backgrounds tend to save more because of education or medical expenditures. Dynan et al. (2004) shows that rich people tend to save more. They used the model of life cycle income with a bequest motive and control variables such as education and future earnings based on age groups. They have tested that a strong positive relationship between current income and saving rates across all income groups (Dynan, et al. 2004), meaning that saving rates tend to increase across different quantiles of income distribution. In addition, the sameness of the essay with the one written by Yin et al. is that they all use the average household income as dependent variable, education, age, and other personal characters as independent variable to get residuals. More residuals mean more risks.

In terms of labour migration, Todaro (1969) researched on the innovation of the model which aimed to formulate the relations between rural-urban migration and income differentials. By examining the probability of finding jobs in urban parts by a rural migrant and the expected income of rural and urban areas, its main conclusion is that when the differentials of rural-urban real income become bigger, more migrants will flow into urban areas to find jobs. It means that rural citizens in a less developed country will tend to seek job opportunities in urban sectors which provide more higher-paying jobs. When the expected probabilities of seeking urban jobs increase, the migration from rural to urban parts of the country will tend to be increased. However, there is a problem regarding the possibility of finding a job in urban sectors. As the author indicates, there may be transitional costs or risks which may prevent rural migrants from coming into urban sectors to find jobs, or they may start from doing a low-paying job for a while and then try to emerge in the other sectors later. In short, the incentives from rationally securing a job in urban sectors even in the long run attract more workers migrant from rural to urban areas.

### III. DATA

The name of the datasets is called The Longitudinal Survey on Rural Urban Migration in China from the Chinese Household Income Project. It was collected by domestic and foreign scholars which have working in the field of economics and statistics since the last century. Most of the datasets includes the detailed information of urban, rural, and migrant individuals and households, recording their economics situations, personal characteristics, family conditions and personal information of these individuals and families. In the end of 1980s, with the rapid development of urbanization and marketization, the issue concerned with income inequality emerged as well. The project also includes five different years of investigation which are respectively 1998, 1995, 2002, 2007 and 2013. It tries to cover most of the basic and detailed information of thousands of households and individuals located in rural and urban areas of China; however, one drawback of these datasets is that they contain many missing values. Although many statistical information came from the official organization of statistics of China, much information derived from the survey is missing because the size of the sample group is too large.

The data used in this essay is from 2007. The reason of selecting this year is that although there is many detailed information about rural and urban households, some of the data cannot convert the code required by the programming app to use. There is also detailed information about migrant individuals and migrant households. The datasets including those information are relevant for me to check and compare the information collected by the rural households. Nonetheless, due to that fact that this essay tries to discover the relationship between the households located in rural parts and their economic situations of saving and consuming, so this essay does not use the migrant information provided by the datasets, rather, it uses the datasets of the information relevant to the rural households and individuals. The variables used in this data involve the personal information, economic situations, and family information of different households and individuals came from a bunch of different provinces. It includes thousands of observations, however, after cleaning the data, a large proportion of data which does not contain any formation has been removed.

   The variables are the education level regarding different households and individuals, whether or not they buy insurance in that year, their health conditions, their martial status and the history of going to different places to work. Because this essay aims to examine the conditions regarding rural citizens working in rural areas and the saving rate, the dependent variable therefore is the household saving rate. According to Yin et.al (2019), there are three methods of calculating the dependent variable which is the saving rate in the context of examining the relations between rural households who have migrants working in urban areas and the saving rates of these rural households. One is to use (total family income-total family expenditure)/ total family income, the other one is more specific to the scenario. It considers that family expenditure may relates to the fee of education and the number of children attending schools, and the medical expenses cannot be estimated, so it takes the method of (total family income- total family expenditure + the expenditure of education and the medical expense)/total family income. As for the third method, Yin et.al (2019) uses the method defined by Chamon and Prasad (2010) to use Ln (family total income/family total expenditure). This paper uses the first way of calculation. However, because of the variables provided by the 2007 dataset define the net income as the combination of different net incomes earned from different sectors including business, financial services, wage, extra, given that there are no other variables indicated as total income in the survey regarding the income and consumption of rural households, this essay, when calculating the total income, defined the total income by adding up the total expenditure summed up by different components of consumption and total net income. Hence, the formula of calculating the household saving rates is (total expenditure + total net income – total expenditure)/ (total expenditure+ total net income). In addition, one thing that needs to be explained here is that there are a lot of missing values in the large dataset. Given that many economic conditions of households are not recorded by this dataset, the missing values have been replaced by zero in terms of preventing the problem of being omitted due to little variance.

Regarding the independent variables, migration has been selected as the independent variable. It comes from the question asked in the survey which is whether or not you worked as a migrant worker in other areas. Due to the lack of information and many missing values in this dataset, the information of many explanatory variables has been replaced by zero in order to do regressions and prevent being omitted. However, this may cause problems that the variance among these variables could be adjusted to surreal conditions.

## IV. MODEL

The model used in this essay is

$$savingrate(i) = \alpha + \beta migartion(i) + \gamma X(i) + \mu(i) \quad (1)$$

From this model, dependent variable represents that the saving rates of a specific household, the explanatory variable is the phenomenon of migration of this household, meaning that someone from this household went to migrant to work in the year. X includes several control variables, and the error term at the end.

There are several terms that are necessary to be explained here. The first one is the issue related to endogeneity. According to Yin et al. (2019), this linear model could have the issues of endogeneity due to the reverse causality and omitted variables. The relation between flexible labor migration and the household saving rates can causality reverse, meaning that the low income resulted in less developed areas could result in the low saving rates. Hence, the family members who are capable of finding jobs in urban sectors are more willing to migrant to urban areas to work. The probabilities of migrating from rural areas to urban sectors are thus higher than working in rural areas domestically. Apart from that, the model could also exist sample sections bias. It is obvious that different jobs have different characters. Some of the jobs require workers especially working as migrants to change the location of their jobs. Thus, it will be better to choose the data which records the information of rural worker from the same level of communities and having other similarities regarding economic level and job characters. In the essay, the method of instrument variable has been adopted to tackle the problem of endogeneity, along with the method of Durbin-Wu-Hausman to test it. In the end, regarding of the robustness of the estimation, the essay adopted the LTZ regression method brought up by Conley et al. (2012) to estimate.

## V. RESULTS

1. The result of descriptive statistics:

| Variable | Obs | Mean | Std. Dev. | Min | Max |
| --- | --- | --- | --- | --- | --- |
| savingr | 46769 | .094 | .813 | -164.673 | 8.881 |
| migration | 46769 | .352 | .965 | 0 | 3 |
| married | 46769 | .409 | .492 | 0 | 1 |
| withinsurance | 46769 | .637 | .513 | 0 | 3 |
| goodhealth | 46769 | .65 | .477 | 0 | 1 |
| badhealth | 46769 | .068 | .402 | 0 | 3 |
| stoc | 46769 | .858 | 3.52 | 0 | 26 |
| numbchi | 46769 | .905 | 1.295 | 0 | 12 |

Figure1: **Descriptive Statistics**

This is the descriptive statistics of different variables which are the dependent variable the household saving rate, migration derived from the question regarding if one of the members of the household went to other places to work in 2007, marital status, insurance, health conditions, the variable stoc represents the education level which is from senior high school to college and the number of children. As explained before, the number of observations is very large. However, this dataset contains many missing values. After cleaning the data, the values replaced the missing values might affect the quality and conciseness of the dataset.

  From the summary, people can see that given the sample is quite large, there exists much imbalance in the data of the variables. According to the mean, in the rural areas, the household saving rates are 9.4 percent, given that the maximum and minimum number of saving rates, it shows that many households have the possibility of owning debts. In terms of migration, the rate of households containing the history of migration as a worker is 35 percent. Due to the description of the question regarding migration, it states that the time period of being a migrant worker must exceed three months, meaning that there could be more households who patriciate in the movement. In the sample, the percentage of households who are married is 40.9 percent which is close to the half of the observations. In terms of the insurance covered by the local governments, business insurance or other. The sample households having different kinds of insurance are over 50 percent, meaning that they all get basic medial care even if they migrant to other places to work. In terms of health conditions, over 65 percent of people have good health, and only 6.8 percent of people having bad health, meaning that the general medical conditions of rural areas are good. In terms of education, the percent of people having finished the "nine-year compulsory education" and high school are 85.8 percent. In addition, people can also see that the maximum year is 26 and the minimum year is 0, meaning that there is possibility that the sample has been affected by missing values; however, the general education level of the sample rural households are good. In terms of the number of children the households have, the maximum number is 12 and the minimum is 0. In fact, when calculating the percent of having children of households, the number of having no children is zero, meaning that the households at least have one child. The data here is adjusted by replacing missing values with zero. However, people can still see that the percent of having children is over 90 percent. Besides, according to the maximum number of children and the variance, though the age of these households is not included, it indicates that many heads of these households in the sample are old people. Before 1980s, most people resided in the less developed rural areas, having many children is common in order to increase the labor to support the family.

In brief, from the descriptive table, people can see that even though many observations are affected by the missing values, it shows that the general condition of the remaining households. In the sample, most of the households residing in different rural places across China have generally good health conditions and basic medical care. In addition, most of them have at least finished the "nine-year compulsory education", indicating that they have acquired a certain level of knowledge to increase the possibility of pursuing different kinds of jobs. In terms of the economic conditions, from the saving rate and the range people can see that many households have debts, along with other expenditures.

2. The result of OLS regression

| savingr | Coefficient | Robust std. err. | t | P>\|t\| | [95% conf. interval] | |
|---:|---:|---:|---:|---:|---:|---:|
| migration | -.0003104 | .0000478 | -6.50 | 0.000 | -.0004041 | -.0002168 |
| married | -.0017618 | .0002686 | -6.56 | 0.000 | -.0022883 | -.0012353 |
| withinsurance | -.0060284 | .0005297 | -11.38 | 0.000 | -.0070667 | -.0049901 |
| goodhealth | -.2835846 | .0112463 | -25.22 | 0.000 | -.3056275 | -.2615418 |
| badhealth | -.116649 | .0046779 | -24.94 | 0.000 | -.1258176 | -.1074803 |
| stoc | -.0000382 | .000015 | -2.54 | 0.011 | -.0000677 | -8.74e-06 |
| numbchi | -.0001895 | .0001157 | -1.64 | 0.101 | -.0004162 | .0000372 |
| _cons | .290746 | .0115143 | 25.25 | 0.000 | .2681777 | .3133143 |

Figure2: OLS regression

This graph shows the correlation between the dependent variable saving rate, independent variables including the explanatory variable which is migration. After controlling for the variables regarding household information and economic conditions, according to the p-values, all the variables, except the number of children are statically significant because their p-values are less than 0.05 significance level. For the variable numbchi indicated the number of children, it is not statistically significant because its p-value is more than 0.05 significance level.

From the coefficients, surprisingly, many literatures regarding the same topic testifies that migration positively contribute to the households saving rate. From the coefficient get from here, it negatively contributes to the household saving rates, meaning that when the rate of migrant labor workers increases, the household saving rates tend to decrease. In terms of marital status, it shows that married people contribute to less household saving rates, meaning that married people may take care of their families, spend more to take care of the children, or are not willing to migrant to seek jobs. These reasons could potentially contribute to the decreasing household saving rates. In terms of the insurance rate, it shows that people having insurance rates tend to contribute negatively to the saving rates. It may show that many of the households buy their own insurance like business insurance, meaning that it could result in the reduction of the income. In terms of the health conditions, compared to people of good health status, people having bad health contribute less to the reduction of the household saving rates, meaning that people need to save more to prepare the money for the sickness. Alternatively, it might be the case that people in good health condition can be able to spend more in the society. In terms of the education level, people having junior high degrees or having obtained college degree are not willing to save more because they usually live in a stable life and want to consume more. In terms of the number of children, presumably, if the scale of the family is large, parents are willing to spend more to take care of their children. Combined with the descriptive statistics of the number of children, it has already shown that the sample households have more than one child, meaning that the variable makes sense to have the negative relation with the dependent variable.

In brief, the conventional Ordinary Least Square regression shows that these variables all more or less contribute negatively to the dependent variable which is the household saving rates. From which, the reverse relation between the variable people having good health and the dependent variable is the strongest. The independent variable migration surprisingly contributes negatively to the dependent variable, meaning that migration from rural places to other places to work could contribute negatively to the household saving rates.

However, these results could be affected by the issue of reverse causality or omitted variable, causing the problem of endogeneity. For example, it is possible that due to the low level of household saving rates, people with good health might want to migrant to work. Thus, the essay uses the method of Instrument Variable in order to tackle the problem of endogeneity.

```
----------------------------
                    (1)
                 savingr
----------------------------
migration         -0.0742***
                  (0.000)

married            0.0142***
```

```
                         (0.000)

numbchi                 -0.0128***
                         (0.000)

stoc                    0.00193***
                         (0.000)

withinsura~e           -0.00729***
                         (0.000)

badhealth              -0.00591***
                         (0.000)

_cons                    0.502***
                         (0.000)
---------------------------
N                         46769
---------------------------
p-values in parentheses
* p<0.05, ** p<0.01, *** p<0.001

Figure3: the result from using Instrument Variable
```

According to the first stage regression and its results[4] including p-value and F-value, the regression show that the results are statistically significant at 0.05 significance level. As the variable goodhealth as an instrument, the difference from the OLS regression is that the household save rates are positively related with the education level and married people, meaning that people who are married and having more years of education could be more willing to migrant to work and increase the household saving rates without considering the effect of health conditions.

## VI. CONCLUSIONS

The findings of the results of analyzing the sample are different from the assumption which is that migration could be able to increase the household save rates because they can send money back to the households, rather, it contributes negatively to the household saving rates, meaning that when a household has more people migrant to work, the household save rates might be lower. The other ones are that household save rates are negatively related with the variables including married people, health conditions, insurance, education level and number of children. The results could suggest that people obtained high-level of education and having a family potentially seek a stable life. Given the summary of the results, people can also find that the households of the sample have high rates of having babies. High percentage of having acquired public insurance and education means that the welfare in the rural areas is not bad. However, due to the sample contained many missing values and the issues with endogeneity, even though that the test of the method of Instrument Variable has been placed, the results presented in here may not be very precise because of the sample bias. In the future, it may be better to choose a better dataset and use different methods to test the issue of endogeneity and different possibilities of the results.

---

[4] Check Data Appendix. (p.18)

## VII. DATA APPENDIX

First-stage regressions

Number of obs    =    46,769

F(  6,  46762) =    1162.28
Prob > F        =     0.0000
R-squared       =     0.1050
Adj R-squared   =     0.1049
Root MSE        =     0.9133

```
                    Robust
migration |    Coef.   Std. Err.    t    P>|t|    [95% Conf. Interval]
-----------------------------------------------------------------------
married     |  .195431   .0158195   12.35  0.000    .1644247   .2264374
numbchi     | -.172189   .0045556  -37.80  0.000   -.1811181  -.1632599
stoc        |  .0261057  .0018185   14.36  0.000    .0225415   .0296699
withinsura~e| -.0838831  .019449    -4.31  0.000   -.1220035  -.0457627
badhealth   |  .2001735  .0119746   16.72  0.000    .1767031   .2236439
goodhealth  |  .6804387  .0204841   33.22  0.000    .6402895   .7205879
cons        |  .0027661  .0004136    6.69  0.000    .0019556   .0035767
```

Tests of endogeneity
Ho: variables are exogenous

Robust score chi2(1) = 568.664 (p = 0.0000)
Robust regression F (1,46761) = 635.841 (p = 0.0000)